\documentstyle[preprint,eqsecnum,aps]{revtex}
\input epsf.sty
\begin{document}

\title{Detecting Event Horizons and Stationary Surfaces}
\author{Richard G. Gass, F. Paul Esposito, L.C.R. Wijewardhana and Louis Witten }
\address{Department of Physics, University of Cincinnati,
Cincinnati,  OH 45220-0011} 
\maketitle

\begin{abstract}               
We have investigated the behavior of three  curvature invariants for Schwarzschild, Reissner-Nordstr{\o}m, Kerr,
and Kerr-Newman black holes. We have also studied these invariants for a Schwarzschild-de Sitter space-time, the $\gamma$ metric, and for a 2+1 charged dimensional black hole.
The invariants are $I_{1}=R_{\alpha\beta\mu\nu;\lambda}R^{\alpha\beta\mu\nu;\lambda}$ , $I_{2}=R_{\mu\nu;\lambda} R^{\mu\nu;\lambda}$,
and $I_{3}=C_{\alpha\beta\mu\nu;\lambda}C^{\alpha\beta\mu\nu;\lambda}$. For all but the Kerr-Newman case these invariants serve
as either horizon or stationary surface detectors. The Kerr-Newman case is more complicated. 
We show that $I_{1}$ vanishs on the horizon in any space-time with a Schwarzschild like metric.

\end{abstract}

\section{Introduction}
An outstanding challenge in the calculation of gravitational collapse to black hole configurations is the identification of emerging black holes in the data
on space-like slices of the space time \cite{S,B,T1}. Since the definition of event horizons involves knowledge of the future evolution of the data,
the computational situation is not as clear-cut as one would like. In fact, the computational burden is shifted towards the identification of apparent horizons, as discussed in the previously cited references. Hoping to alleviate some of this burden, we have tried to develop criteria, involving the curvature tensor and its
derivatives, which would indicate the existence of black holes. These criteria would, ideally, be complementary to those  involving marginally outer-trapped surfaces.
These criteria also would indicate a trend in the data towards the formation of black holes. This hope has been fulfilled only partially. 
We take the opportunity to correct a result which has previously appeared in the literature.

It has long been believed that an observer falling toward a black hole cannot tell
when he or she has crossed the event horizon. This belief is based on the fact that
the horizon is a global property of the geometry. The invariant

\begin{equation}
I_{1}=R_{\alpha\beta\mu\nu;\lambda}R^{\alpha\beta\mu\nu;\lambda}
\end{equation} 

\noindent involves a covariant derivative of the curvature tensor and thus may be used to probe a neighborhood of the  event horizon.
 This means that it has a
 chance of functioning as a horizon
detector. This fact was first noticed by Karlhede, Lindstr\"{o}m, and {\AA}man \cite{K} who
 calculated $I_{1}$ for
 the Schwarzschild, Reissner-Nordstr{\o}m, Taub-NUT, and Kerr solutions. Karlhede, Lindstr\"{o}m and {\AA}man found that $I_{1}$ vanished on the event horizon for 
 the Schwarzschild, Reissner-Nordstr{\o}m, and Taub-NUT solutions, and that it vanished on the stationary surface for the Kerr solution.
 Our results differ from those
 of Karlhede {\it at el} by a sign and due to our sign convention which is that of Misner, Thorne, and Wheeler \cite{MTW}.  $I_{1}$ has also been investigated by Tammelo and Kask \cite{T}.
  To our knowledge, the invariants 
  
\begin{equation}
I_{2}=R_{\mu\nu;\lambda} R^{\mu\nu;\lambda}
\end{equation}
 and 
 
 \begin{equation}
 I_{3}=C_{\alpha\beta\mu\nu;\lambda}C^{\alpha\beta\mu\nu;\lambda}
 \end{equation}
 
\noindent have not been previously investigated. The curvature invariants are, in principle, locally measurable quantities \cite{K,T,K2}, and thus could allow the local detection of an event horizon or a stationary surface.

Using {\it Mathematica} and MathTensor, we have calculated $I_{1}$, $I_{2}$, and $I_{3}$ for the Kerr-Newman geometry and for the special
cases of Kerr, Reissner-Nordstr{\o}m, and Schwarzschild black holes. We have also computed $I_{1}$ for a Schwarzschild-de Sitter
space-time, for the $\gamma$ metric \cite{E}, and for a charged 2+1 black hole space-time.  
On a 300 MegHz PowerMac G3 the calculations take several hours of run time for the longest (Kerr-Newman) cases.
For vacuum space-times, $I_{1}=I_{3}$ and, of course, $I_{2}=0$. We conjecture that $I_{1}$ serves as an event horizon detector for any static space-time where
the horizon is regular. If the horizon is singular, it is not clear how to distingish the horizon from any other type of singularity.

In addition to being intrinsically interesting horizon detection is of considerable interest
in numerical relativity. The ablility to track the causal structure of a space-time is believed to be a crucial step
in numerically evolving black hole space-times \cite{S}. For this reason considerable effort has gone into constructing apparent horizon
 finders in numerical relativity \cite{B,T1}. The invariants discussed here maybe useful in that regard.
 
\section{Static and Stationary Space-Times}
We have investigated the behaviour of $I_{1}$ for both four dimensional space-times and a 2+1 black hole space-time.

\subsection{Four Dimensional Space-Times}
\subsubsection{Schwarzschild black holes}
We begin by discussing the Schwarzschild solution since it is the simplest case.

\noindent For the Schwarzschild solution, the metric is 
\begin{equation}
ds^2=-(1-2M/r) dt^2 +(1-2M/r)^{-1} dr^2 +r^2(d\theta^2+\sin^2 \theta d\phi^2),
\end{equation}
and we find that \cite{K}
\begin{equation}I_{1}= -{\frac{720\,{M^2}\,\left( 2\,M - r \right) }{{r^9}}}.\end{equation}

It is clear from equation (2.2) that the invariant 
is positive outside the event horizon, goes to zero on the event horizon, and is negative inside the event horizon.
The invariant has a maximum at $r=9M/4$, and  the only zeros of the invariant are at infinity and on the event horizon.
An in-falling observer can thus detect the presence of the event horizon by monitoring $I_{1}$. The in-falling observer may, for instance, measure
 the variation in oscillation frequencies from the equation of geodesic deviation \cite{T}.
 An in-falling observer can not save him or her self by turning back when the
invariant  vanishes for
  by that time it is too late! However, the invariant does provide a warning and the observer could turn back when the maximum of the invariant is reached.
\subsubsection{Reissner-Nordstr{\o}m black holes}
The metric for the Reissner-Nordstr{\o}m solution is
\begin{equation}
ds^2=-(1-2M/r+Q^2/r^2) dt^2 +(1-2M/r +Q^2/r^2)^{-1} dr^2 +r^2(d\theta^2+\sin^2 \theta d\phi^2).
\end{equation}
The Reissner-Nordstr{\o}m case is somewhat more complicated, but the results for $I_{1}$ are essentially the same as those for the Schwarzschild solution:
  $I_{1}$ vanishes 
only at $r=\infty$, on the horizon $r_{+}$, and on the null surface $r_{-}$. 
There is a positive extremal value of $I_{1}$ just outside of $r_{+}$.  The full invariant is
   
   \begin{equation} I_{1}= {\frac{16\,\left( 76\,{Q^4} - 108\,M\,{Q^2}\,r + 
       45\,{M^2}\,{r^2} \right) \,
     \left( {Q^2} + r\,\left( -2\,M + r \right)  \right) }
     {{r^{12}}}},\end{equation}
     
     \noindent and we see that $I_{1}=0$ at $ r= \infty $.
The vanishing of $I_{1}$ can be reduced to finding the roots where the numerator of equation 2.4 vanishes.

  There are two  finite complex roots and two finite real roots. The two real roots are the null surface  $r_{-}=M - {\sqrt{{M^2} - {Q^2}}}$ and the event horizon
  $r_{+}=M + {\sqrt{{M^2} - {Q^2}}}$.  This result was first obtained by Karlhede, Lindstr\"{o}m and {\AA}man \cite{K}, although there is a typographical error
    in their result for $I_{1}$ for the Reissner-Nordstr{\o}m case. Figure one shows a  typical plot of $I_{1}$ as a function of $r$ in the vicinity of $r_{+}$ and
figure 
   two shows a  typical plot of $I_{1}$ as a function of r in the vicinity of $r_{-}$. We notice that $I_{1}$ is positive outside
$r_{+}$, 
   negative in the region between $r_{+}$, and $r_{-}$ and positive inside $r_{-}$. 
   The complex roots occur at $r={\frac{2\,\left( 9\,{Q^2} - i\,{\sqrt{14}}\,{Q^2} \right) }
   {15\,M}}$ and $r={\frac{2\,\left( 9\,{Q^2} + i\,{\sqrt{14}}\,{Q^2} \right) }
   {15\,M}}$.

   For the Reissner-Nordstr{\o}m case, $I_{2}$  
   
   \begin{equation} I_{2}= {\frac{80\,{Q^4}\,\left( {Q^2} + 
       r\,\left( -2\,M + r \right)  \right) }{{r^{12}}}} .\end{equation}

    \noindent We observe that, like $I_{1}$,  $I_{2}$ vanishes at  $r_{+}$ and $r_{-}$, but that unlike $I_{1}$, $I_{2}$ has no
       complex roots. Figure three shows $I_{2}$ in the region of the horizon.
       
       The invariant $I_{3}$ is given by

 \begin{equation} I_{3} ={\frac{48\,\left( {Q^2} - 2\,M\,r + {r^2} \right) \,
     \left( 22\,{Q^4} - 36\,M\,{Q^2}\,r + 
       15\,{M^2}\,{r^2} \right) }{{r^{12}}}}.\end{equation}
       
  \noindent  Like $I_{1}$ and $I_{2}$, $I_{3}$ vanishes at $r_{+}$ and $r_{-}$. In addition $I_{3}$ has
   two complex roots which lie at $r={\frac{18\,{Q^2} - i\,{\sqrt{6}}\,{Q^2}}{15\,M}}$ and $r={\frac{18\,{Q^2} + i\,{\sqrt{6}}\,{Q^2}}{15\,M}}.$
   The complex roots of $I_{1}$ and $I_{3}$ are not the same but near the horizon $I_{1}$ and $I_{3}$ are numerically close in the sense that $ \vert (I_{1}-I_{3})/I_{1} \vert \ll 1$ . Figure four 
   shows the difference between $I_{1}$ and $I_{3}$ in the vicinity of the horizon for a particular choice of parameters.
   
 For an extremal black hole $Q=M$ the two roots at $r_{+}$ and $r_{-}$ coalesce into a double root at $r=M$. If $Q^2 > M^2$, then $I_{1}$, $I_{2}$,
  and $I_{3}$ have no real roots.

  \subsubsection{ Schwarzschild-de Sitter spacetimes}
 For the Schwarzschild-de Sitter space-time the metric is
 
 \begin{equation}
ds^2=-(1-2M/r+\Lambda r^2/3) dt^2 +(1-2M/r+\Lambda r^2/3)^{-1} dr^2 +r^2(d\theta^2+\sin^2 \theta d\phi^2),
\end{equation}

\noindent and  

\begin{equation} I_{1}={\frac{240\,{M^2}\,\left( -6\,M + 3\,r - {r^3}\,\Lambda  \right)
       }{{r^9}}}. \end{equation}

\noindent This invariant is zero when $1- 2M/r -\Lambda r^{2}/3=0$, which is the event horizon. Thus,  $I_{1}$ seems 
to serve as a cosmological event horizon detector.

\subsubsection{Space-Times with Schwarzschild like metrics}

We call a Schwarzschild like metric one with the line element 
\begin{equation}
ds^{2} = -A(r) dt^2 + A(r)^{-1}dr^2 +r^{2} d \theta ^{2} + r^{2} \sin^{2} \theta d \phi ^{2}.
\end{equation}

\noindent where $A(r)$ is an arbitrary function of r.
For this metric we can calculate $I_{1}$ for a general $A(r)$ and we find that

\begin{eqnarray}
I_{1} & = &  A(r)\left[\,\left( 32 + 32\,{{A(r)}^2} + 
       16\,{r^2}\,{{A'(r)}^2} - 
       32\,A(r)\,\left( 2 + r\,A'(r) \right) \right. \right. \nonumber \\
  & & \left. \left.    + 8\,{r^4}\,{{A''(r)}^2} - 
       16\,r\,A'(r)\,\left( -2 + {r^2}\,A''(r) \right)  + 
       {r^6}\,{{A'''(r)}^2} \right) \right] /r^{6}, 
\end{eqnarray} 
\noindent where the primes indicate differentiation with respect to $r$. We notice that $I_{1}$ vanishes when $A(r)$, the norm of the time-like Killing
vector, vanishes. In addition $I_{1}$ will vanish when

\begin{eqnarray}
32 + 32\,{{A(r)}^2} + 16\,{r^2}\,{{A'(r)}^2} - 
  32\,A(r)\,\left( 2 + r\,A'(r) \right) \nonumber \\ + \,
  8\,{r^4}\,{{A''(r)}^2} - 
  16\,r\,A'(r)\,\left( -2 + {r^2}\,A''(r) \right)  + 
  {r^6}\,{{A'''(r)}^2}=0.
\end{eqnarray} 
With the appropriate choices for $A(r)$, $I_{1}$ reduces to either the Schwarzschild or the Reissner-Nordstr{\o}m result.
We have also calculated the invariants $I_{2}$ and $I_{3}$ for this space-time and find that
\begin{eqnarray}
I_{2}& = & A(r)\left[\,\left( 24 + 24\,{{A(r)}^2} + 
       4\,{r^2}\,{{A'(r)}^2} + 24\,{r^2}\,A''(r) + 
       10\,{r^4}\,{{A''(r)}^2} - \right. \right. \nonumber \\
      & & \left. \left. 24\,A(r)\,\left( 2 + {r^2}\,A''(r) \right)  + 
       4\,{r^5}\,A''(r)\,A^{(3)}(r) + 
       {r^6}\,{{A^{'''}(r)}^2} - \right. \right. \nonumber \\
      & & \left. \left. 4\,{r^3}\,A'(r)\,\left( 2\,A''(r) + r\,A^{'''}(r)
           \right)  \right) \right]/2\,{r^6}
\end{eqnarray}
and 
\begin{eqnarray}
I_{3}& = &  A(r)\left[\,\left( 40 + 40\,{{A(r)}^2} + 
       40\,{r^2}\,{{A'(r)}^2} - 40\,{r^2}\,A''(r) + 
       10\,{r^4}\,{{A''(r)}^2} + 8\,{r^3}\,A^{'''}(r) - 
       4\,{r^5}\,A''(r)\,A^{'''}(r) +  \right. \right. \nonumber \\
       & & \left. \left. {r^6}\,{{A^{'''}(r)}^2} +
    8\,r\,A'(r)\,\left( 10 - 5\,{r^2}\,A''(r) + 
          {r^3}\,A^{'''}(r) \right)  - 
       8\,A(r)\,\left( 10 + 10\,r\,A'(r) - \right. \right. \right. \nonumber \\
        & & \left. \left. \left.  5\,{r^2}\,A''(r) + {r^3}\,A^{'''}(r) \right) 
        \right) \right]/3\,r^{6}.
\end{eqnarray}
If the space-time is a vacuum space-time then the field equations imply 
\begin{equation}
{{A'(r)} = {{\frac{1 - A(r)}{r}}}}
\end{equation}
and $I_{2}=0$, while 
\begin{equation}
I_{1}=I_{3}={\frac{180\,{{\left( -1 + A(r) \right) }^2}\,A(r)}{{r^6}}}.
\end{equation}

\subsubsection{ Kerr Black Holes}
The metric for the Kerr solution is
\begin{equation}
ds^2=\rho^2 \left(\frac{dr^2}{\Delta}+d\theta^2 \right)+(r^2+a^2) \sin^2 \theta d \phi^2-dt^2+\frac{2 M r}{\rho^2}\left( a\sin^2 \theta d \phi -dt\right)^2,
\end{equation}
where $\rho^2(r,\theta) = r^2 + a^2 \cos^2\theta$ and $\Delta (r) = r^2 -2 M r+a^2$.
For Kerr black holes the only invariant of interest is $I_{1}$, and we find the remarkably simple result that
\begin{eqnarray}
I_{1}&=& 368640\,{M^2}\,\left( -2\,M\,r + {r^2} + 
       {a^2}\,{{\cos (\theta )}^2} \right) \, \times \nonumber \\
     && \frac{\left( {r^8} - 28\,{a^2}\,{r^6}\,{{\cos (\theta )}^2} + 
       70\,{a^4}\,{r^4}\,{{\cos (\theta )}^4} - 
       28\,{a^6}\,{r^2}\,{{\cos (\theta )}^6} + 
       {a^8}\,{{\cos (\theta )}^8} \right)} {{{\left( {a^
           2} + 2\,{r^2} + {a^2}\,\cos (2\,\theta ) \right) }
       ^9}}.
\end{eqnarray}

\noindent The form of our result is different from the one given by Karlhede {\it at el} \cite{K} but the two agree up to the previously mentioned sign convention.

For the Kerr solution, the invariant $I_{1}$ has 10 roots. Two of these involve the mass of the hole and are at the inner and outer stationary limits given by 
$r_{\pm}=M\pm \sqrt{M^{2}-a^{2} \cos^2 \theta}$. The other eight
roots depend only on $a$. The fact that $I_{1}$ vanishes at the stationary limit rather than the horizon is not surprising since the norm of the 
Killing vector $\bf{\frac{\partial}{\partial t}}$ vanishes on the stationary surface. The Killing vector
 ${\bf\frac{\partial}{\partial t}} + \Omega_{H} {\bf\frac{\partial}{\partial \phi}}$  ,where $\Omega_{H}=\frac{a}{r_{+}^{2}+a^{2}}$ is tangent
to the null geodesic generators of the horizon. \cite{W}. 

The eight Kerr roots that depend only on $a$  may be expressed in a  simple form and are given in Table 1.
 We observe that the roots are paired; for every root at a positive value of $r$ there is a root at the corresponding negative value of $r$. Each root in the
 left column of Table 1 is paired with the corresponding root in the right column. The existence of roots for negative values of $r$ is perhaps
 not surprising since the maximally extended Kerr solution  includes both positive and negative values of $r$ \cite{H}. 
 As  $a \to 0 $, the mass independent roots coalesce to zero. If $a^2>M^2$ then the two mass dependent roots become complex.

The zeros of $I_{1}$ are shown in figure 5. The behavior of $I_{1}$ in the vicinity of the horizon is shown in figure 6.

\subsubsection{Kerr-Newman Black Holes}
 The most complicated case we consider is the Kerr-Newman black hole. The metric for the Kerr-Newman solution is
 
\begin{eqnarray}
ds^2 & = &-(\Delta/\rho^2)\left(dt - a \sin^2 \theta d\phi \right)^{2} +(\sin^2 \theta / \rho^2 )\left[ \left(r^2+a^2 \right) d \phi -a dt \right]^{2} \nonumber \\
&& +(\rho^2/ \Delta) dr^2 +\rho^2 d\theta^2,
\end{eqnarray}
where $\rho^2(r,\theta) = r^2 + a^2 \cos^2\theta$ and $\Delta (r) = r^2 -2 M r+a^2+Q^2$.

 For the Kerr-Newman solution, all three invariants are distinct and in general do not vanish
 on either the horizon or on the stationary surface. We find that
\small{ \begin{eqnarray}
 I_{1}&=& \left[8192\,\left( {r^6}\,
        \left( 76\,{Q^6} - 260\,M\,{Q^4}\,r + 
          261\,{M^2}\,{Q^2}\,{r^2} + 76\,{Q^4}\,{r^2} - 
          90\,{M^3}\,{r^3} \right. \right. \right.  \nonumber \\
          & & \left. \left. \left.  - 108\,M\,{Q^2}\,{r^3} + 
          45\,{M^2}\,{r^4} + 
          {a^2}\,\left( 44\,{Q^4} - 36\,M\,{Q^2}\,r \right) 
           \right)  + {a^2}\,{r^4}\, \right. \right. \times \nonumber \\
       & & \left. \left. \left( -744\,{Q^6} + 3540\,M\,{Q^4}\,r - 
          5364\,{M^2}\,{Q^2}\,{r^2} - 712\,{Q^4}\,{r^2} + 
          2520\,{M^3}\,{r^3} + 1980\,M\,{Q^2}\,{r^3} - \right. \right. \right. \nonumber \\
          & & \left. \left. \left.1215\,{M^2}\,{r^4} + 
          {a^2}\,\left( -156\,{Q^4} + 324\,M\,{Q^2}\,r
             \right)  \right) \,{{\cos (\theta )}^2} + 
       6\,{a^4}\,{r^2}\,\left( 98\,{Q^6} - 
          706\,M\,{Q^4}\,r + 1545\,{M^2}\,{Q^2}\,{r^2} \right. \right. \right. \nonumber \\ 
          & & \left. \left. \left. - 1050\,{M^3}\,{r^3} - 222\,M\,{Q^2}\,{r^3} + 
          315\,{M^2}\,{r^4} + 
          {a^2}\,\left( -26\,{Q^4} + 30\,M\,{Q^2}\,r \right) 
           \right) \,{{\cos (\theta )}^4} + \right. \right. \nonumber \\
      & & \left. \left. 2\,{a^6}\,\left( -16\,{Q^6} + 302\,M\,{Q^4}\,r - 
          1170\,{M^2}\,{Q^2}\,{r^2} + 356\,{Q^4}\,{r^2} + 
          1260\,{M^3}\,{r^3} - 1350\,M\,{Q^2}\,{r^3} + 
          945\,{M^2}\,{r^4} + \right. \right. \right. \nonumber \\
         & & \left. \left. \left. 2\,{a^2}\,\left( 11\,{Q^4} - 45\,M\,{Q^2}\,r
             \right)  \right) \,{{\cos (\theta )}^6} - 
       {a^8}\,\left( 76\,{Q^4} + 90\,{M^3}\,r - 
          720\,M\,{Q^2}\,r - 
          45\,{M^2}\,\left( {Q^2} - 27\,{r^2} \right) 
          \right) \,{{\cos (\theta )}^8} + \right. \right. \nonumber \\
      & & \left. \left. 45\,{a^{10}}\,{M^2}\,{{\cos (\theta )}^{10}} \right) \right]
     /{{\left( {a^2} + 2\,{r^2} + 
         {a^2}\,\cos (2\,\theta) \right) }^9},
 \end{eqnarray}}
 \small{
 \begin{eqnarray}
 I_{2}&=& {\frac{-2048\,{Q^4}\,\left( -\left( {r^2}\,
          \left( 4\,{a^2} + 
            5\,\left( {Q^2} - 2\,M\,r + {r^2} \right) 
            \right)  \right)  - 
       {a^2}\,\left( 4\,{a^2} - {Q^2} + 2\,M\,r \right) \,
        {{\cos (\theta )}^2} + 5\,{a^4}\,{{\cos (\theta )}^4}
        \right) }{{{\left( {a^2} + 2\,{r^2} + 
         {a^2}\,\cos (2\,\theta ) \right) }^7}}},
 \end{eqnarray}} and 
\small{ \begin{eqnarray}
 I_{3}& = &
\left[ 24576\,\left( {r^6}\,
        \left( 22\,{Q^6} - 80\,M\,{Q^4}\,r + 
          87\,{M^2}\,{Q^2}\,{r^2} + 22\,{Q^4}\,{r^2} - 
          30\,{M^3}\,{r^3} - 36\,M\,{Q^2}\,{r^3} + 
          15\,{M^2}\,{r^4} + \right. \right. \right. \nonumber \\
       & & \left. \left. \left.  12\,{a^2}\,\left( {Q^4} - M\,{Q^2}\,r \right) 
           \right)  + {a^2}\,{r^4}\,
        \left( -254\,{Q^6} + 1192\,M\,{Q^4}\,r - 
          1788\,{M^2}\,{Q^2}\,{r^2} - 244\,{Q^4}\,{r^2} + \right. \right. \right. \nonumber \\
        & & \left. \left. \left. 840\,{M^3}\,{r^3} + 660\,M\,{Q^2}\,{r^3} - 
          405\,{M^2}\,{r^4} + 
          {a^2}\,\left( -60\,{Q^4} + 108\,M\,{Q^2}\,r \right)
              \right) \,{{\cos (\theta )}^2} + \right. \right. \nonumber \\
      & & \left. \left. 2\,{a^4}\,{r^2}\,\left( 97\,{Q^6} - 
          704\,M\,{Q^4}\,r + 1545\,{M^2}\,{Q^2}\,{r^2} - 
          1050\,{M^3}\,{r^3} - 222\,M\,{Q^2}\,{r^3} + 
          315\,{M^2}\,{r^4} - \right. \right. \right. \nonumber \\
         & & \left. \left. \left.  30\,{a^2}\,\left( {Q^4} - M\,{Q^2}\,r \right) 
           \right) \,{{\cos (\theta )}^4} + 
       2\,{a^6}\,\left( -5\,{Q^6} + 100\,M\,{Q^4}\,r - 
          390\,{M^2}\,{Q^2}\,{r^2} + 122\,{Q^4}\,{r^2} +   \right. \right. \right. \nonumber \\
         & & \left. \left. \left.   420\,{M^3}\,{r^3} -
450\,M\,{Q^2}\,{r^3} + 
          315\,{M^2}\,{r^4} + 
          6\,{a^2}\,\left( {Q^4} - 5\,M\,{Q^2}\,r \right) 
           \right) \,{{\cos (\theta )}^6} - 
       {a^8}\,\left( 22\,{Q^4} + \right. \right. \right. \nonumber \\
		& & \left. \left. \left. 30\,{M^3}\,r - 
          240\,M\,{Q^2}\,r - 
          15\,{M^2}\,\left( {Q^2} - 27\,{r^2} \right) 
          \right) \,{{\cos (\theta )}^8} + \right. \right. \nonumber \\
      & & \left. \left. 15\,{a^{10}}\,{M^2}\,{{\cos (\theta )}^{10}} \right) ) \right] 
     /{{\left( {a^2} + 2\,{r^2} + {a^2}\,\cos (2\,\theta ) \right) }^9}. 
 \end{eqnarray}}
 
 Although $I_{1}$ and $I_{3}$ are distinct, the surfaces on which they vanish have, for typical values of the hole parameters, similar shapes. Both of these invariants have 
 10 roots, of which two depend on the mass of the hole. The other eight roots depend only on the charge and angular momentum. 
  The surfaces on which $I_{1}$ vanishes are shown in figures 7, 8, 9, and 10. The roots of $I_{3}$ are similar. As $a \to 0$, the roots 
  move toward the Reissner-Nordstr{\o}m case. The physical significance of the mass independent roots is unclear. As $a \to 0$, the two mass dependent roots move to the inner and outer horizons and the mass independent roots
  coalesce into the two complex roots of the Reissner-Nordstr{\o}m case. 
  This is shown in figure 11.  
  
  The invariant $I_{2}$ has four roots which are shown in figure 12 for a hole with $M=2, Q=1$, and  $a=1$. We notice that the outer root lies in the ergosphere, and that the
  second root weaves in and out of the inner horizon. The two remaining roots lie inside the inner horizon. In the limit $a \to 0$, the two inner roots
  coalesce to zero.
  The roots of $I_{2}$ for an extreme Kerr-Newman hole are shown in figure 13. Although it is not obvious from the figure, $I_{2}$ also vanishes on the event horizon 
  at $\theta =0$ and $\pi$.
   The two mass dependent roots remain real for some values of $\theta$ even for the ultra-extreme case where $Q^2+a^2 >M^2$.
  In general none of the invariants vanishes on the horizon nor on the stationary surface, except at $\theta =0$. Some of the roots of $I_{1}$, $I_{2}$, and $I_{3}$
  are complex for some values of $a, Q$ and $\theta$, as they must be, since both $I_{1}$ and $I_{2}$ have two imaginary roots for the Reissner-Nordstr{\o}m case.
  
 \subsubsection{The $\gamma$ metric}
 
 A static, axially symmetric, asymptotically flat vacuum space-time is provided by the $\gamma$ metric \cite{E}, a two parameter Weyl solution. The $\gamma$ metric 
 serves as an example of a space-time where the horizon, depending on the 
 values of the parameters $M$ and $\gamma$, may  be regular or not. The $\gamma$ metric is
 \begin{eqnarray}
 ds^{2}& = & - 
     {\left( 1 - \frac{2\,M}{r} \right) }^{\gamma }{dt}^2 
   +  {\left( 1 - \frac{2\,M}{r} \right) }^
     {-\gamma } \left[\left( \frac{r^{2}- 2M r}{r^{2} -2 M r +
      M \sin \theta ^{2}} \right)^{\gamma^{2}-1} dr^2 +  \right. \nonumber \\
    && \left.  \frac{\left(r^2 - 2M r\right)^{\gamma^2}}{\left(r^2 - 2 M r +M^2 \sin \theta^{2} \right) ^{\gamma^2-1}} d \theta ^2 +
     (r^2-2 M r) \sin \theta ^{2} d\phi^2 \right],
     \end{eqnarray}
     
    \noindent and corresponds to a total mass of $\gamma M$.
 The $\gamma$ metric reduces to the Schwarzschild metric when $\gamma=1$. While the complete analysis of the $\gamma$ metric is rather complicated, we  mention only a few of the features which have
 a bearing on the present calculations. For $0<\gamma <2$, $\gamma \neq 1$  the null surface $r=2M$ is singular 
 along the symmetry axis $\theta =0$ or $\pi$. For $\gamma \geq 2$ the null surface $r=2M$ is regular along the symmetry axis.
 This can be seen by looking at $I_{0}=R_{\alpha \beta \mu \nu} R^{\alpha \beta \mu \nu}$.  
 We have  computed  both $I_{0}$  and $I_{1}$ for general $\theta$ but the results are long. 
 Along the symmetry axis we find
 
 \begin{equation}
 I_{0}={\frac{48\,{M^2}\,{{\left( 1 - {\frac{2\,M}{r}} \right) }^
       {2\,\gamma }}\,{{\gamma }^2}\,
     {{\left( M - r + M\,\gamma  \right) }^2}}{{r^4}\,
     {{\left( -2\,M + r \right) }^4}}}
     \end{equation}
     
     \noindent and
    
 \begin{equation}
     I_{1}={\frac{80\,{M^2}\,{{\left( 1 - {\frac{2\,M}{r}} \right) }^
       {3\,\gamma }}\,{{\gamma }^2}\,
     {{\left( 3\,{r^2} - 
          6\,M\,r\,\left( 1 + \gamma  \right)  + 
          2\,{M^2}\,\left( 2 + 3\,\gamma  + 
             {{\gamma }^2} \right)  \right) }^2}}{{r^6}\,
     {{\left( -2\,M + r \right) }^6}}},
     \end{equation}
     
     \noindent which reduces to the Schwarzschild result if $\gamma =1$. $I_{1}$ vanishes on the horizon if $\gamma \geq 2$. There are also two zeros of $I_{1}$
     found symmetrically about $r=M(1+\gamma)$. In short $I_{1}$, is singular
     on the horizon when the horizon is singular and vanishes on the  horizon when the horizon is regular. The belief that Weyl solutions are important
     in collapse situations hinges on the hope that reasonable interior solutions can be found for them.

 \subsection{2+1 Black Holes}
We have also computed $I_{1}$, $I_{2}$, and $I_{3}$ for the BTZ 2+1 dimensional black hole. 
The charged BTZ black hole found by Ba\~{n}ados, Teitelboim, and Zanelli \cite{B2} has a metric given by

\begin{equation}
ds^{2}=-\left(\frac{r^{2}}{\sigma^{2}}-M +Q^{2} \ln r \right) dt^{2} + \left(\frac{r^{2}}{\sigma^{2}}-M +Q^{2} \ln r \right)^{-1} dr^{2} +r^{2} d \phi^{2}.
\end{equation}

\noindent For the charged BTZ black hole
\begin{equation}
I_{0}={\frac{12\,{r^4} + 4\,{Q^2}\,{r^2}\,{\sigma^2} + 
     3\,{Q^4}\,{\sigma^4}}{{r^4}\,{\sigma^4}}}.
\end{equation}
The scalar curvature is 
\begin{equation}
R= \frac{-(Q^2-6 r^2/\sigma^2)}{r^2},
\end{equation}
\noindent and the invariant $I_{1}$ is 
\begin{equation}
I_{1}= {\frac{20\,{Q^4}\,\left( -M + {(r/\sigma)^2} + {Q^2}\,\log (r) \right)
       }{{r^6}}},
\end{equation}
which is seen by inspection to vanish on the horizon.
The invariant $I_{2}$ is found to be
\begin{equation}
I_{2}={\frac{6\,{Q^4}\,\left( {r^2} - M\,{\sigma^2} + 
       {Q^2}\,{\sigma^2}\,\log (r) \right) }{{r^6}\,{\sigma^2}}}.
\end{equation}

\noindent The space-time is Weyl flat, so $I_{3}=0$.

Since the geometrical and physical interpretation of the curvature invariants is unclear it may be useful to
rewrite $I_{2}$ in terms of $F_{\mu \nu}$.  Using the Einstein equations, we find that $I_{2}$ may be written
in terms of invariants involving $F_{\mu \nu}$ and $R$. It is straightfoward to
show that 
\begin{equation}
I_{2} = 4 \left( I_{EM1} -\frac{5}{16}I_{EM2} \right) + \frac{3}{4} R_{; \lambda} R^{; \lambda},
\end{equation}
\noindent where 
\begin{equation}
I_{EM1}= \left( F_{\mu \beta} F_{\nu}^{\beta} \right)_{; \lambda}\left( F^{\mu}_{ \beta} F^{\nu \beta} \right)^{; \lambda}, 
\end{equation}
\noindent and 
\begin{equation}
I_{EM2} = \left(F_{\alpha \beta} F^{\alpha \beta} \right)_{; \lambda} \left(F_{\alpha \beta} F^{\alpha \beta} \right)^{; \lambda}.
\end{equation}
We find that 
\begin{equation}
I_{EM1}={\frac{2\,{Q^4}\,\left( {r^2} - M\,{\sigma^2} + 
       {Q^2}\,{\sigma^2}\,\log (r) \right) }{{r^6}\,{\sigma^2}}},
\end{equation}

\begin{equation}
I_{EM2}={\frac{4\,{Q^4}\,\left( {r^2} - M\,{\sigma^2} + 
       {Q^2}\,{\sigma^2}\,\log (r) \right) }{{r^6}\,{\sigma^2}}},
\end{equation}
\noindent and

\begin{equation}
R_{; \lambda} R^{; \lambda}= {\frac{4\,{Q^4}\,\left( {r^2} - M\,{\sigma^2} + 
       {Q^2}\,{\sigma^2}\,\log (r) \right) }{{r^6}\,{\sigma^2}}}.
\end{equation}
Both the electromagnetic invariants and the scalar curvature invariant vanish on the horizon, and are in fact proportional to $I_{2}$.

The uncharged black hole space-time has constant scalar curvature and is, for our proposes, uninteresting since $I_{1}$ is identically zero.

\section{Conclusions}
The invariants $I_{1}$, $I_{2}$, and $I_{3}$, which are locally measurable, serve as horizon or stationary surface detectors for a variety of space-times but not for Kerr-Newman black holes.
We have shown that $I_{1}$ serves as a horizon detector for any Schwarzschild like space-time and
 conjecture that $I_{1}$ serves as a horizon detector for regular horizons in any static axially symmetric space-time.

It  is not known if an 
invariant can be found that will vanish on the horizon in the Kerr-Newman case.

A problem of considerable current interest in  numerical general relativity is the detection of emerging horizons in collapse problems.
The invariants discussed here, particularly $I_{1}$ and $I_{2}$, may be  useful tools for the numerical detection of such emerging horizons.

\section{Acknowledgments}
One of the authors (R.G.) thanks Daniel Lichtblau for a useful conversation. {\it Mathematica} is a registered trademark of Wolfram Research Inc.
MathTensor is a registered trademark of MathSoft Inc. L.C.R.W. and L.W. acknowledge the partial support of the U.S. Department of Energy under contract number DOE - FG02-84ER40153.

 \begin{table}
\small{$\begin{array}{cc}
r=\left( 1 + {\sqrt{2}} + {\sqrt{4 + 2\,{\sqrt{2}}}} \right) \,a\,\cos (\theta ) &  r=-\left( \left( 1 +
{\sqrt{2}} + {\sqrt{4 + 2\,{\sqrt{2}}}} \right) \,a\,\cos (\theta ) \right) \\
r=\left( 1 + {\sqrt{2}} - {\sqrt{4 + 2\,{\sqrt{2}}}} \right) \,a\,\cos (\theta ) & r= \left( -1 -
{\sqrt{2}} + {\sqrt{4 + 2\,{\sqrt{2}}}} \right) \,a\,\cos (\theta ) \\
r=\left( -1 + {\sqrt{2}} \right) \,a\,\cos (\theta ) + 
  {\sqrt{4 - 2\,{\sqrt{2}}}}\,{\sqrt{{a^2}\,{{\cos (\theta )}^2}}} & r= -\left( \left( -1 + {\sqrt{2}} \right) \,a\,\cos (\theta ) \right)  - 
  {\sqrt{4 - 2\,{\sqrt{2}}}}\,{\sqrt{{a^2}\,{{\cos (\theta )}^2}}}\\
r=-\left( \left( -1 + {\sqrt{2}} \right) \,a\,\cos (\theta ) \right)  + 
  {\sqrt{4 - 2\,{\sqrt{2}}}}\,{\sqrt{{a^2}\,{{\cos (\theta )}^2}}} & r=\left( -1 + {\sqrt{2}} \right)
\,a\,\cos (\theta ) - 
  {\sqrt{4 - 2\,{\sqrt{2}}}}\,{\sqrt{{a^2}\,{{\cos (\theta )}^2}}}
\end{array}$}
\caption{The eight roots of $ I_{1}=0 $ that are independent of the mass of the hole for the Kerr solution.}
\end{table}
  
\begin{figure}
\epsfxsize= 3in
\centerline{ \epsffile{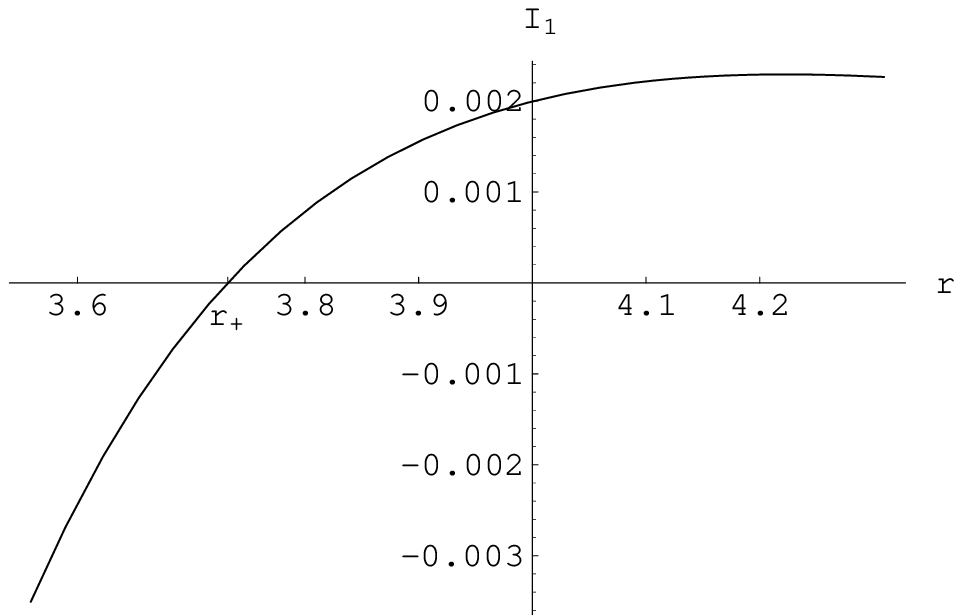}}
\centerline{\caption{$I_{1}$ for the Reissner-Nordstr{\o}m solution as a function of $r$ for the case $M=2,Q=1$ in the vicinity of $r_{+}$}}
\end{figure} 
  
\begin{figure}
\epsfxsize= 3in
\centerline{ \epsffile{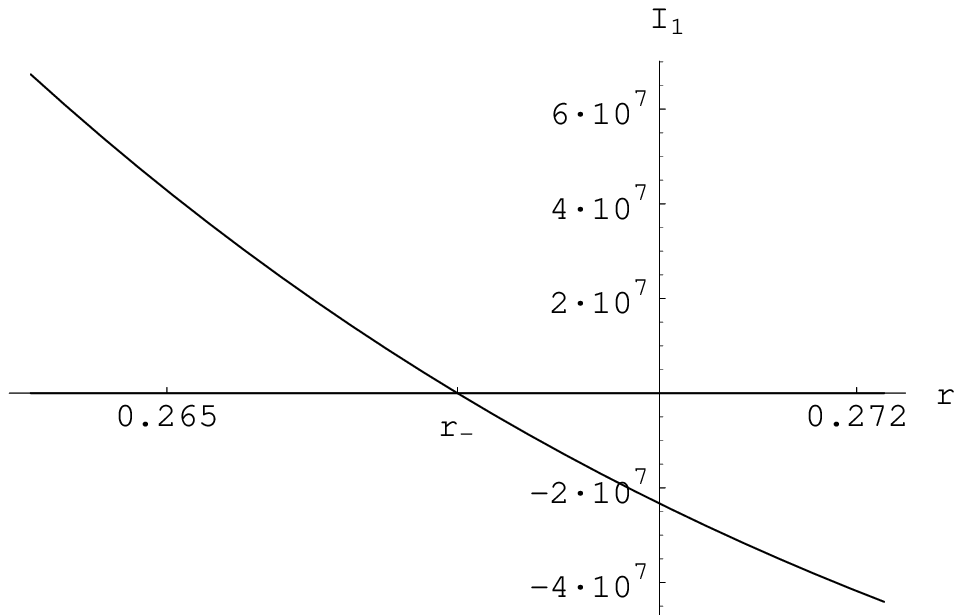}}
\centerline{\caption{$I_{1}$ for the Reissner-Nordstr{\o}m solution as a function of $r$ for the case $M=2,Q=1$ in the vicinity of $r_{-}$}}
\end{figure}

\begin{figure}
\epsfxsize= 3in
\centerline{ \epsffile{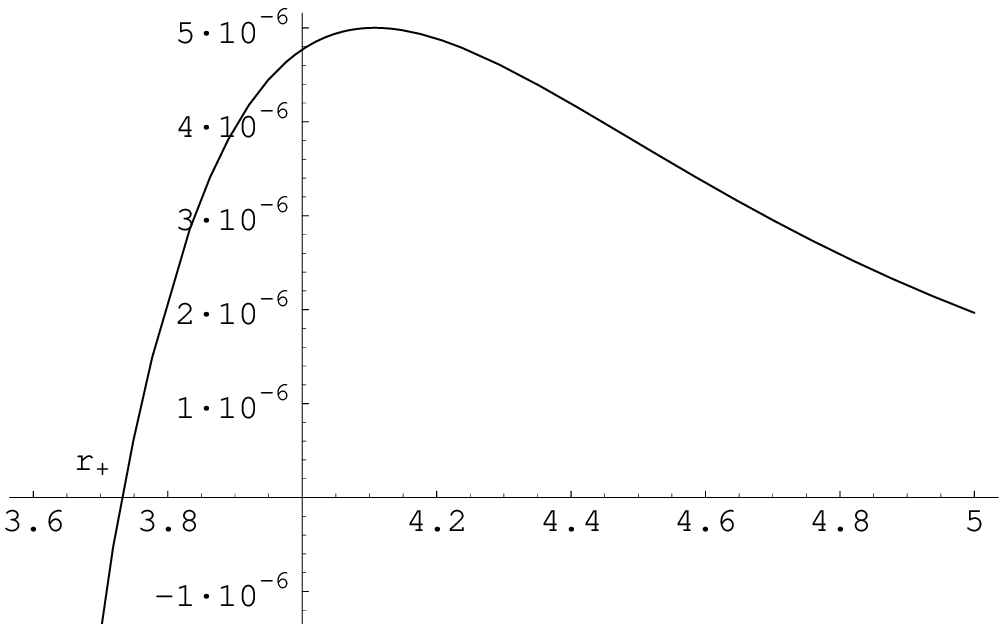}} 
\centerline{\caption{$I_{2}$ for the Reissner-Nordstr{\o}m solution as a function of $r$ for the case $M=2,Q=1$ in the vicinity of $r_{+}$}}
\end{figure} 

\begin{figure}
\epsfxsize= 3in
\centerline{ \epsffile{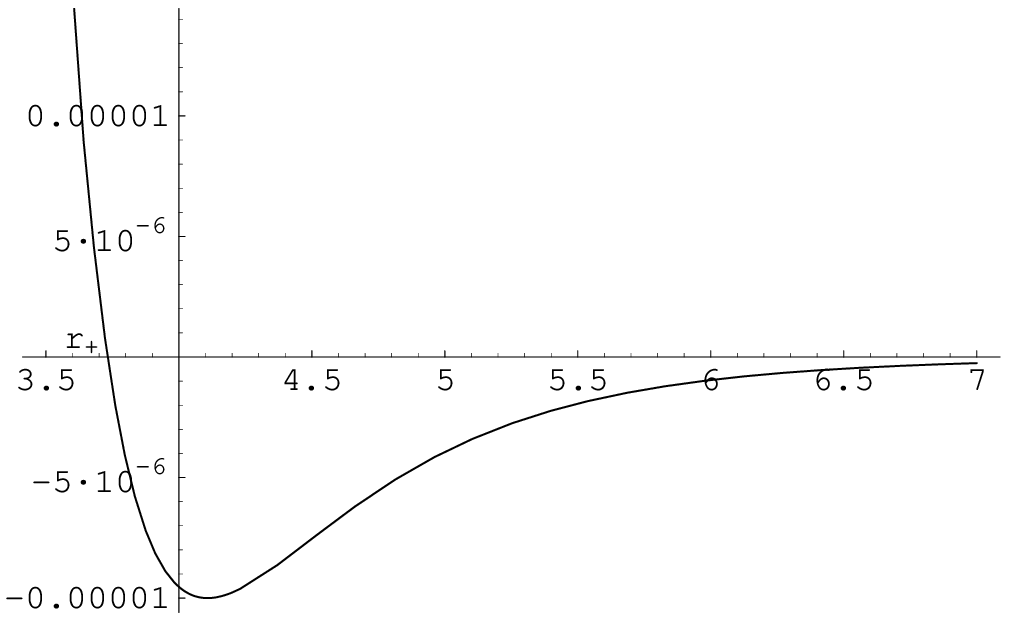}}
\centerline{\caption{$I_{3}-I_{1}$ for the Reissner-Nordstr{\o}m solution as a function of $r$ for the case $M=2,Q=1$ in the vicinity of $r_{+}$}}
\end{figure} 

\begin{figure}
\epsfxsize= 5in
\centerline{ \epsffile{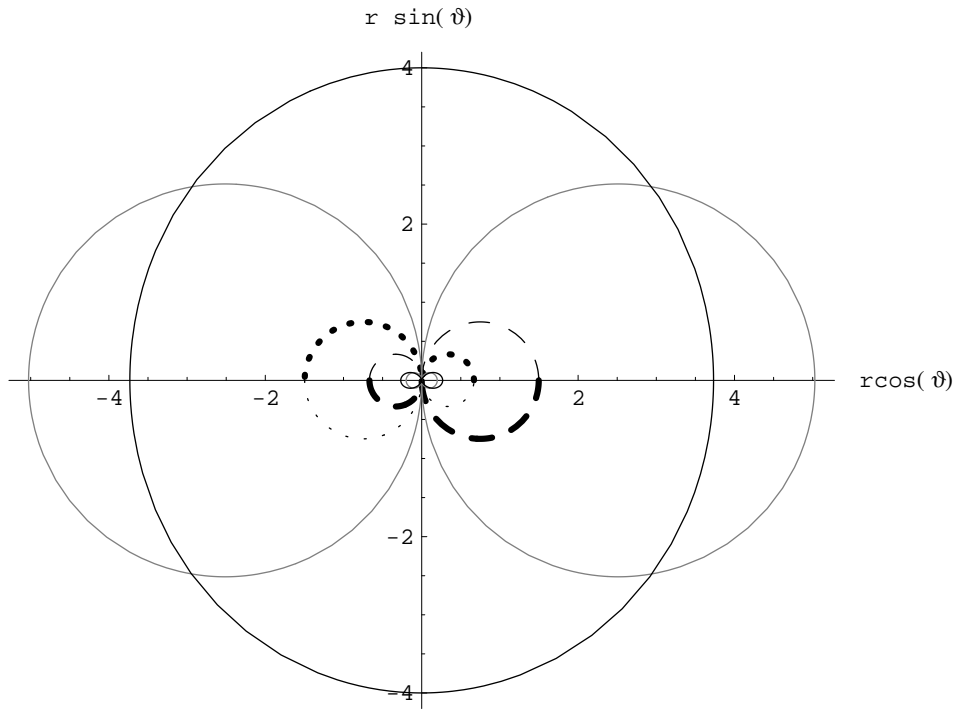}}
\centerline{\caption{ The roots of $I_{1}$ for the Kerr metric. The inner and outer stationary surfaces are shown in black. There is a root
on each stationary surface. Four other roots
 are shown in grey. The remaining four roots are shown as dashed lines with a different dashing used for each root. Two of the roots are inside the inner stationary surface and are thus hard to see. All the roots except the two which lie on the stationary surfaces
  depend only on a.}}
\end{figure}

\begin{figure}
\epsfxsize= 3in
\centerline{ \epsffile{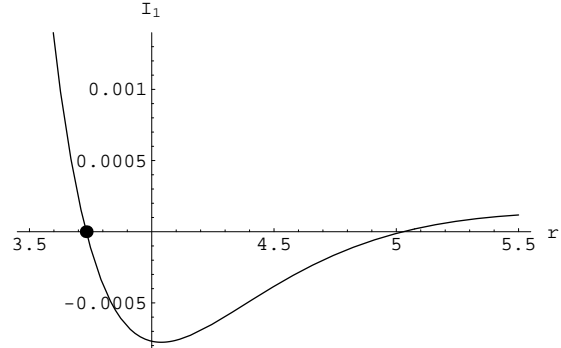}}
\centerline{\caption{$I_{1}$ or the Kerr metric as a function of $r$ for the case $M=2,a=1$. 
The approach is along the $\theta = 0$ axis and the stationary limit is indicated by the black dot.}}
\end{figure} 
 
 \begin{figure}
\epsfxsize= 5in
\centerline{ \epsffile{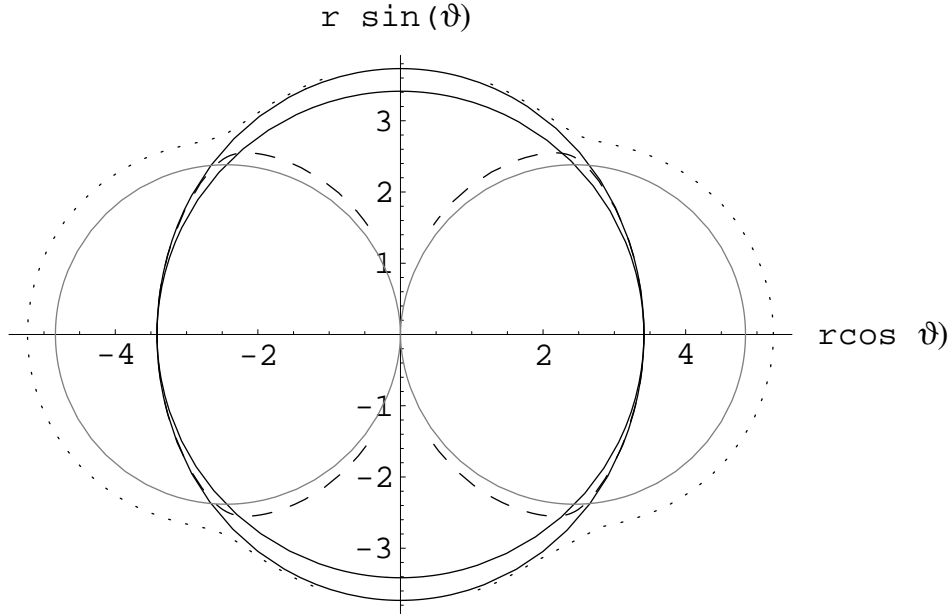}}
\centerline{\caption{ The outer roots of $I_{1}$ for the Kerr-Newman metric with $M=2,a=1,Q=1$. The horizon and outer stationary surfaces are shown in black.
 The three outer roots are shown in gray and as dashed lines.}}
\end{figure} 

\begin{figure}
\epsfxsize= 5in
\centerline{ \epsffile{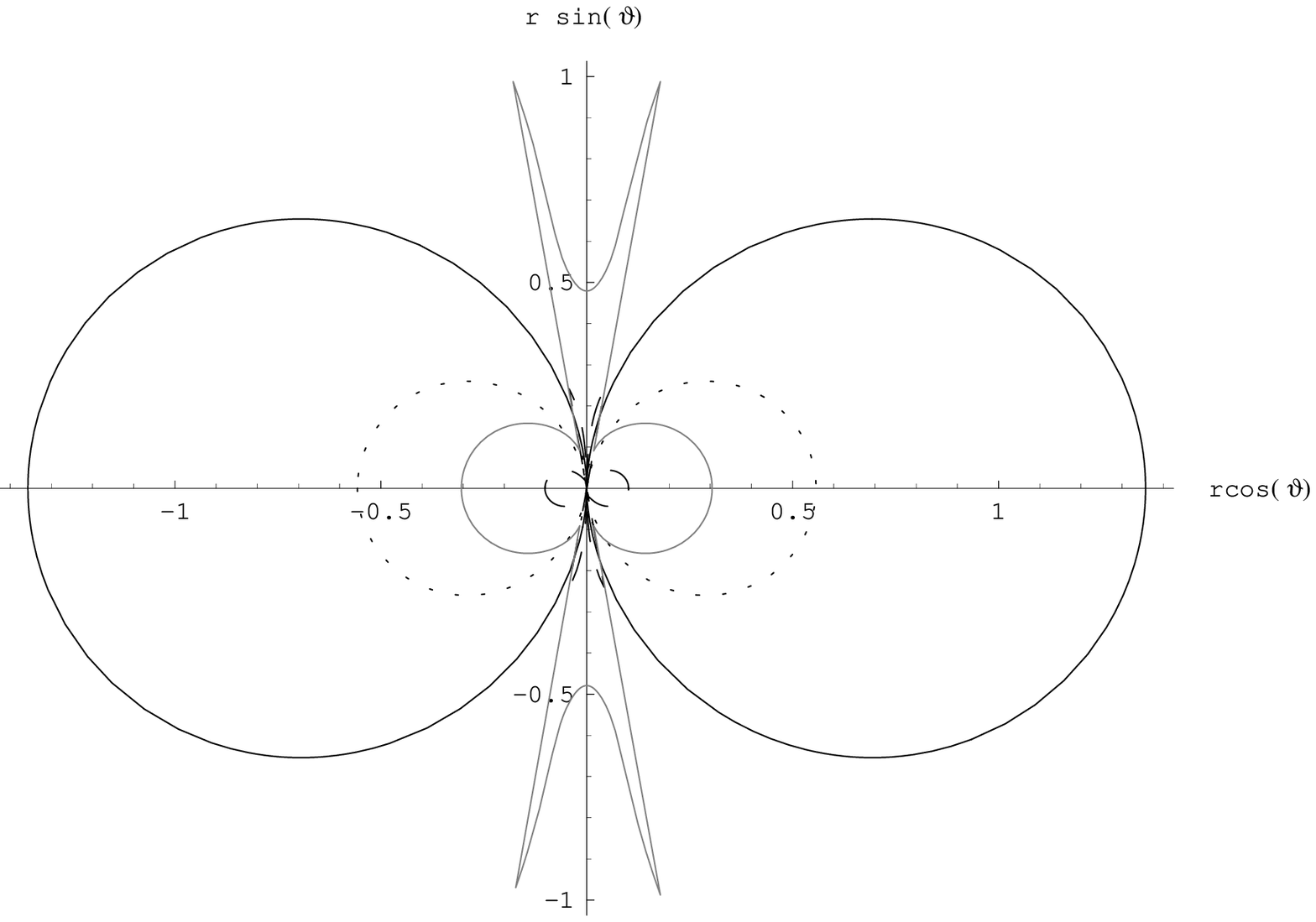}}
\centerline{\caption{ Four of the inner roots of $I_{1}$ for the Kerr-Newman metric with $M=2,a=1,Q=1$.
 A close up view of the inner most root is shown in figure 9. }}
\end{figure} 

\begin{figure}
\epsfxsize= 5in
\centerline{ \epsffile{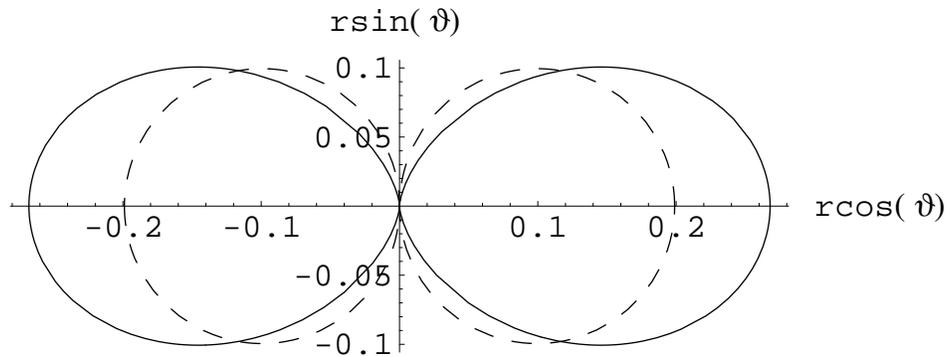}}
\centerline{\caption{ The inner most root (shown with a dashed line) of $I_{1}$ and the inner stationary surface (shown in black)
 for the Kerr-Newman metric with $M=2,a=1,Q=1$.}}
\end{figure}

\begin{figure}
\epsfysize= 5in
\centerline{ \epsffile{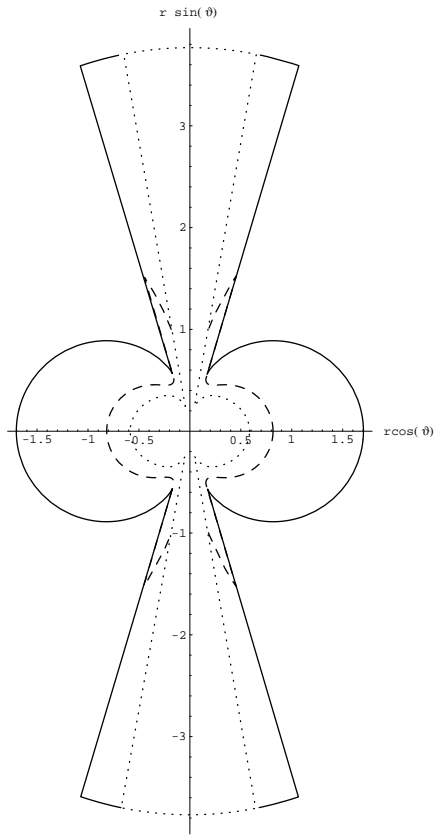}}
\centerline{\caption{ The other three  inner roots of $I_{1}$ for the Kerr-Newman metric
 with $M=2,a=1,Q=1$.
 These roots all lie outside the inner stationary surface.}}
\end{figure} 

\begin{figure}
\epsfxsize= 7in
\centerline{ \epsffile{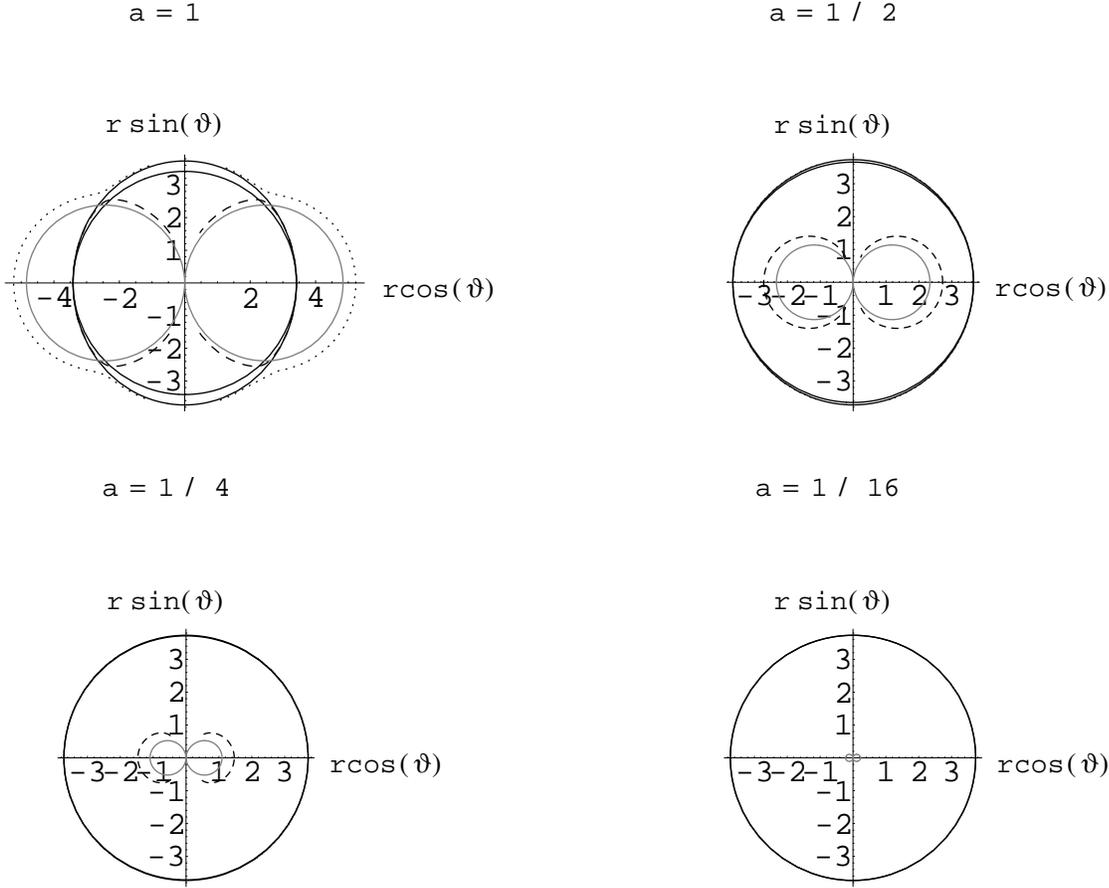}}
\centerline{\caption{ The  three outer roots of $I_{1}$ for the Kerr-Newman metric as $a \to 0$ with $M=2,Q=1$. As the hole spins down the roots 
  move toward the Reissner-Nordstr{\o}m case. The outer horizon and stationary surface are shown in black.
   The two mass dependent roots move to the inner and outer horizons and the mass independent roots
  coalesce into the two complex roots of the Reissner-Nordstr{\o}m case. For reasons of scale only one of the mass independent roots is shown. The root shown with the dashed line becomes complex as $a \to 0$.}}
\end{figure} 



\begin{figure}
\epsfxsize= 6in
\centerline{ \epsffile{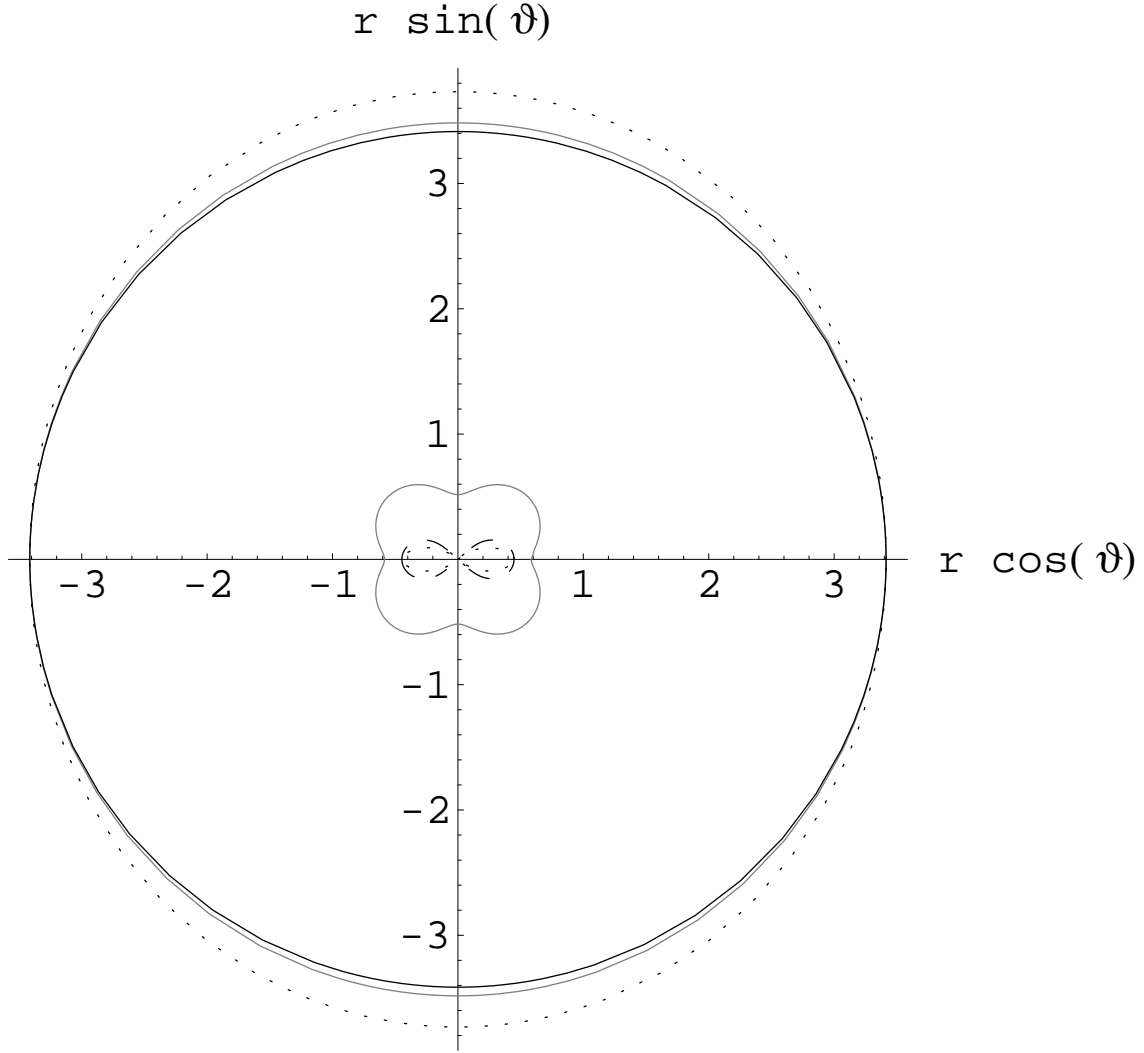}}
\centerline{\caption{ The  four roots of $I_{2}$ for the Kerr-Newman metric for $M=2,a=1,Q=1$. 
The  outer root, shown in gray, is trapped in the ergosphere. The stationary limit is show 
as a dashed line, the event horizon
is shown in black. The inner  gray root weaves in and out of the inner horizon which is not shown.}}
\end{figure} 

\begin{figure}
\epsfxsize= 4in
\centerline{ \epsffile{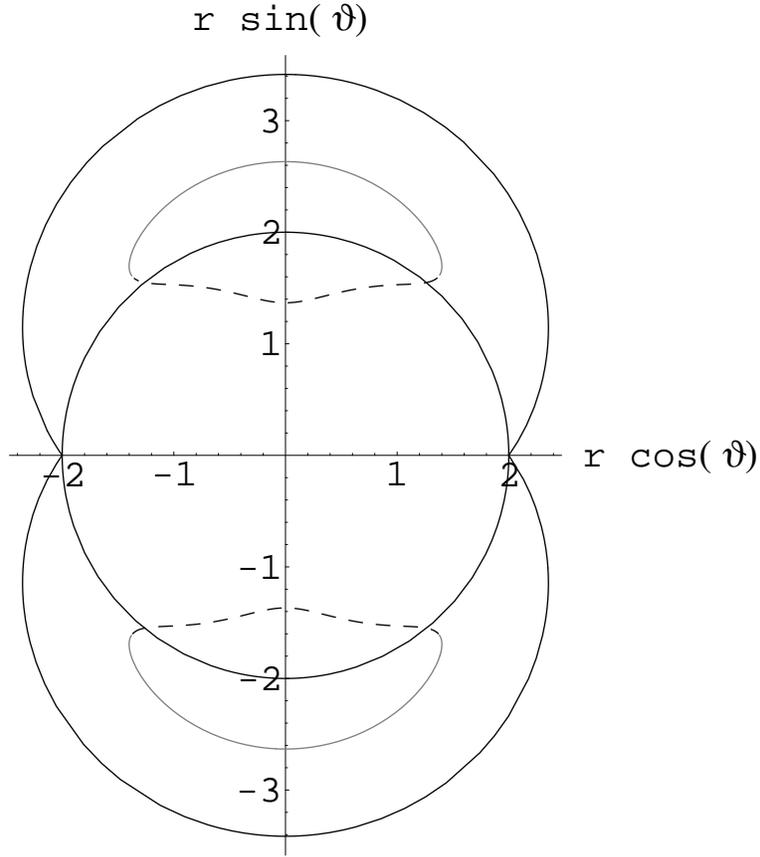}}
\centerline{\caption{ The  two outer roots of $I_{2}$ for the Kerr-Newman metric for an extreme case with $M=2,a= 2^{1/2},Q=2^{1/2}$.
 The two outer roots have joined each other. The light gray line shows the same root as the outer gray line in figure 12.
 The dashed line is same root as the inner gray root in figure 12. The horizon and the stationary surface are shown in black.}}
\end{figure}

\end{document}